\documentclass[preprint,aps,floatfix,superscriptaddress]{revtex4}
\usepackage{graphicx,epsfig,amssymb,amsmath,array}
\usepackage{relsize,placeins,float}
\usepackage[dvipsnames]{xcolor}
\usepackage{blindtext}
\usepackage[english]{babel}
\newcommand{\cfcoh}{coupling--forcing induced coherence}

\begin{document}
\title{Interplay of coupling and forcing: coherence in a nonlinear delayed oscillator}

\author{Mattia Coccolo}
\affiliation{Nonlinear Dynamics, Chaos and Complex Systems Group, Departamento de F\'{i}sica,
Universidad Rey Juan Carlos, Tulip\'{a}n s/n, 28933 M\'{o}stoles, Madrid, Spain}

\author{Miguel A.F. Sanju\'{a}n}
\affiliation{Nonlinear Dynamics, Chaos and Complex Systems Group, Departamento de F\'{i}sica,
Universidad Rey Juan Carlos, Tulip\'{a}n s/n, 28933 M\'{o}stoles, Madrid, Spain}
\affiliation{Royal Academy of Sciences of Spain, Valverde 22, 28004 Madrid, Spain}

\date{\today}

\begin{abstract}
Coupled oscillators often experience two distinct influences: an internal input transmitted through coupling from a driver subsystem, and an external periodic forcing applied directly to the response. Here we ask how these two influences interact and whether their combined action can organize the response dynamics. We show that coupling and forcing can cooperate to produce a \emph{localized} organization in parameter space, where the response displays simultaneously enhanced steady-state amplitudes and increased spectral concentration at the forcing frequency; we term this effect \emph{coupling--forcing induced coherence}. In delay regimes where the driver approaches a steady state, the coupling acts as an effective bias that shifts the response operating point, providing a resonance-like mechanism that helps anchor the onset of the coherent band. When the driver is also externally forced, coupling--forcing induced coherence can coexist with \emph{transmitted resonance} (amplification conveyed from the driver to the response through the coupling pathway). We use frequency- and time-domain diagnostics to identify parameter regions dominated by coupling--forcing induced coherence, by transmitted resonance, or by their combined action.
\end{abstract}

\maketitle

\section{Introduction}

Driven systems play a crucial role in diverse scientific disciplines, ranging from medicine \cite{jiruska, mormann, Rulkov} and physics \cite{jensen, hramov} to communication \cite{Koronovskii, Naderi}, mechanics \cite{defoort}, networks \cite{delellis, zhang}, circuits \cite{Yao}, and engineering \cite{sujith}. They provide insights into collective and emergent behaviors, including phase transitions in ferromagnetic materials investigated through mean-field coupling \cite{Desai, Dawson}. Winfree \cite{Winfree} proposed a foundational model for populations of coupled nonlinear oscillators with globally stable limit cycles, and in neuroscience driven systems have been used to describe brain rhythms and synchronization-related phenomena \cite{Wiener}. Beyond synchronization, however, driven nonlinear systems can also exhibit amplification and organization effects that emerge from the interplay of distinct perturbations.

The study of resonance phenomena has long been central to nonlinear science. Many forms of nonlinear resonance, such as stochastic resonance \cite{Gammaitoni, McDonnell}, chaotic resonance \cite{Zambrano}, vibrational resonance \cite{Landa}, delay-induced resonance \cite{Cantisan, Coccolo}, and Bogdanov--Takens resonance \cite{Coccolo2}, have been explored, each arising from different mechanisms. Resonance plays a fundamental role in various physical systems, from musical instruments and celestial mechanics to engineering applications. At the same time, the broader concept of coherence, where oscillations become more organized or spectrally concentrated under the influence of perturbations, has received increasing attention in fields such as neuroscience, physics, and engineering \cite{Breakspear, Larger, Pikovsky, Boccaletti}.

In previous work~\cite{Coccolo_synch}, we studied a coupling-induced amplification scenario in which a time-delayed driver system acted as the only external influence on a non-delayed response system. In that setting, the response exhibited a resonance-like enhancement generated by the interaction between the coupling input and the response dynamics. In a subsequent study~\cite{Coccolo_synch2}, we added an external periodic
forcing to the driver and found that its resonance-like behavior could be
conveyed to the response through the coupling pathway (transmitted resonance).
Moreover, we showed that introducing delay in the driver can effectively
represent an upstream feedback or processing latency whose effects are conveyed
to the response through the same pathway. For consistency, we adopt the same
configuration here and focus on delay regimes in which the driver dynamics
remain regular. In other delay regions, the response organization is typically
degraded due to irregular driver activity or large-amplitude excursions.

In this work, we extend these investigations by analyzing the response system under two simultaneous external influences: a coupling-mediated perturbation generated by a time-delayed driver system and an independent periodic forcing applied directly to the response. Their cooperation leads to a phenomenon that we term \textit{coupling--forcing induced coherence}. Here, ``coherence'' refers to a localized organization of the response dynamics in parameter space, manifested by enhanced steady-state amplitudes together with spectral concentration at the forcing frequency. While a resonance-like mechanism may contribute to this effect for specific delay values, coherence rather than resonance alone provides the unifying description across the explored parameter regimes. We emphasize that this deterministic mechanism differs from classical coherence resonance, where regular oscillations emerge through noise-induced transitions near an excitability threshold: in our case, coherent organization arises without stochasticity from the nonlinear interplay between periodic forcing and coupling-mediated influence. Moreover, such coherence can coexist with the transmitted resonance observed in coupled systems. The central aim of this work is therefore to disentangle the respective contributions of coupling--forcing induced coherence and transmitted resonance, and to clarify when localized amplification originates from one, the other, or their combined action.

In this context, we distinguish coherence from resonance: while resonance typically denotes amplification at a well-defined frequency, coherence encompasses the broader organization and localization of oscillations, including both amplitude and spectral structure, which may occur with or without strict frequency locking. This perspective better captures the phenomena observed in our system, where enhanced organization may arise in regimes that do not support clear resonant frequency locking.

Various coupling mechanisms have been explored in the literature \cite{Pecora1, Pecora2, Pecora3}, but we adopt the continuous control coupling method \cite{Ding, Kapitaniak} because it provides a direct and analytically tractable representation of the driver system's influence on the response system as a coupling-mediated input. This choice facilitates a systematic study of how coupling and external forcing cooperate or compete in shaping coherent response regimes.

Although our study focuses on a particular nonlinear oscillator configuration, the interaction between periodic forcing and coupling-induced dynamics is a general phenomenon observed in many contexts. For instance, in mechanics, forced and coupled pendula can display amplitude localization and coherence effects \cite{Blekhman}, while in electronics, forced RLC circuits and Josephson junction arrays show coherence amplification under combined coupling and forcing \cite{Pikovsky,Larger}. And in neuroscience, models of coupled neurons demonstrate coherence enhancement in spiking activity due to the interplay of external inputs and synaptic coupling \cite{Breakspear,Ermentrout}.

The mechanism explored here may therefore be relevant in physical settings where a subsystem is simultaneously influenced by an internal coupling signal and an external periodic drive. For example, in mechanical and electromechanical oscillators, nonlinear micro- and nanomechanical resonators can display amplitude control, bistability, and frequency stabilization under external driving and coupling-like actuation~\cite{Mahboob2008,Antonio2012}. Related coupling-mediated signal injection protocols have also been used in experimental oscillator platforms~\cite{Marro}. In electronic systems, driven RLC or Josephson circuits subject to coupling feedback can display analogous coherence amplification, and in neuronal models, the interaction between synaptic input or coupling and rhythmic stimulation can promote coherent oscillatory activity within specific parameter regimes~\cite{Buzsaki2006,Wilson2016}.

Therefore, after introducing the model in Sec.~\ref{Sec:II}, we analyze the coherence phenomenon in Sec.~\ref{Sec:III}. In particular, we examine the induced organization in the response oscillations due to the cooperation between the coupling constant and the external forcing for different values of the delay. Next, we explore how changes in the coupling strength and in the forcing amplitude impact the response dynamics. Finally, concluding remarks are presented in Sec.~\ref{Sec:concl}.

\section{Model}\label{Sec:II}

The unidirectional coupling used in this study follows a general continuous control scheme. We define the driver system as an autonomous nonlinear dynamical system whose state is represented by a vector \( \mathbf{x}_1 \in \mathbb{R}^n \), governed by the set of nonlinear differential equations
\begin{equation}
    \dot{\mathbf{x}}_1 = \mathbf{F}(\mathbf{x}_1).
\end{equation}
Similarly, the response system is described by a state vector \( \mathbf{x}_2 \in \mathbb{R}^d \) evolving according to
\begin{equation}
    \dot{\mathbf{x}}_2 = \mathbf{G}(\mathbf{x}_2).
\end{equation}
When a unidirectional coupling is introduced, the response system is modified as
\begin{equation}
    \dot{\mathbf{x}}_2 = \mathbf{G}(\mathbf{x}_2) + \mathbf{C}\,(\mathbf{x}_1-\mathbf{x}_2),
\end{equation}
where \( \mathbf{C} \) is a coupling matrix whose elements determine the strength and placement of the interaction. Depending on its structure, the coupling can act on a subset of variables or on the whole response state, and it may include cross-coupling and self-coupling contributions.

The term \( \mathbf{C}\,(\mathbf{x}_1-\mathbf{x}_2) \) provides a continuous, state-dependent influence of the driver on the response while preserving the unidirectional character of the interaction. This formulation is widely used in coupled nonlinear oscillators and networked dynamical systems, including studies of synchronization, resonance-like amplification, and control in physics and engineering~\cite{Pecora1,Pecora2,Pecora3,Breakspear,Larger}.

Having introduced the coupling mechanism, we now specify the coupled oscillators
considered in this work. We study a response Duffing oscillator, denoted by
\(x_2\), influenced by a time-delayed driver Duffing oscillator, denoted by
\(x_1\). The coupling mechanism and the dynamical regimes of the delayed driver
have been thoroughly analyzed in previous works~\cite{Coccolo_synch,
Coccolo_synch2}. In addition, the response is subjected to an independent
external periodic forcing, yielding
\begin{align}\label{eq:1}
Driver&\rightarrow \quad  \frac{d^2x_1}{dt^2}+\mu\frac{dx_1}{dt}+\gamma x_1(t-\tau)+\alpha x_1(1-x_1^2)=0,\\
Response&\rightarrow \quad   \frac{d^2x_2}{dt^2}+\mu\frac{dx_2}{dt}+\alpha x_2(1-x_2^2)=C(x_1-x_2)+f\cos(\omega_2 t),
\end{align}

where we fix the parameters \(\mu=0.01\), \(\alpha=-1\), and \(\gamma=-0.5\). The scalar \(C\) is the coupling constant (the only nonzero entry of the continuous control scheme coupling matrix~\cite{Ding,Kapitaniak}) and quantifies the strength of the driver-to-response influence through the diffusive term \(C(x_1-x_2)\). Although this coupling form is often employed in synchronization studies, here it is adopted as a simple and analytically tractable mechanism of cross-influence. Similar diffusive interactions arise in control theory, neural networks, and delayed feedback systems~\cite{Pyragas,Breakspear,Giacomelli}, where the aim is modulation and shaping of activity rather than enforcing phase-locking. In the present setting, the response is simultaneously affected by a coupling-mediated input (generated by the driver dynamics) and by an independent periodic forcing \(f\cos(\omega_2 t)\). 
We use a unidirectional coupling configuration deliberately. The aim of the
present work is not to study mutual synchronization between two interacting
oscillators, but to isolate how a coupling-mediated input generated by an
upstream delayed subsystem cooperates with a direct periodic forcing applied to
the response. If bidirectional coupling were introduced, the response would
feed back into the driver, modifying the driver dynamics itself and mixing the
coupling--forcing mechanism with additional effects such as mutual
synchronization, feedback-induced transitions, and possible changes of the
delay-induced driver regimes. The unidirectional configuration therefore
provides a cleaner causal setting in which the driver acts as a structured
input and the response is the subsystem where the cooperative organization is
measured. Bidirectional coupling is an interesting extension, but it would
constitute a different problem and is beyond the scope of this article.

Our focus here is on the organization that emerges from their combined action. To appreciate the effects of these two influences on the response dynamics, we consider weak dissipation (\(\mu=0.01\)).
Although we adopt a dimensionless Duffing-type formulation, the ingredients required for coupling--forcing induced coherence are common in laboratory settings: a nonlinear resonator (the response) simultaneously driven by (i) an external periodic actuation and (ii) an internally generated coupling-mediated signal that may be delayed or filtered by an upstream subsystem. In this sense, the delay $\tau$ can be interpreted as an effective upstream feedback or processing latency, while the coupling strength $C$ controls the magnitude of the injected internal signal. Therefore, the phenomenon reported here predicts a \emph{localized} region in forcing parameters where the response becomes both strongly amplified and spectrally concentrated at the forcing frequency; this makes the effect directly measurable from time series via standard amplitude and Fourier diagnostics.

The modeling choice of introducing the time delay only in the driver system is grounded in our previous work~\cite{Coccolo_synch}, where we showed that such a configuration can effectively represent a delayed upstream influence on a non-delayed response subsystem: delay-induced regimes develop in the driver and are then conveyed to the response through the coupling pathway, without explicitly introducing a delay in the response equation. In the present model, the delay is therefore not a literal propagation delay in the coupling term \(C(x_1-x_2)\), but an internal feedback/latency time of the driver subsystem that shapes the coupling-mediated signal received by the response. Its role is twofold: it controls the temporal structure of the driver input and it selects the dynamical regimes of the driver that can support organized response behavior.

From a physical perspective, such configurations are common in systems where a response unit is affected by both a direct periodic forcing and an upstream delayed or feedback-mediated input. For instance, in control systems a delayed controller may act on a non-delayed actuator; in neuroscience, delayed synaptic or processing effects may influence local neural dynamics without requiring an intrinsic delay in the postsynaptic unit; and in electromechanical systems, delayed feedback may perturb an otherwise memoryless nonlinear oscillator. In electromechanical/control realizations, a periodically forced nonlinear resonator, such as a beam, MEMS device, or pendulum, may receive an additional delayed feedback-like signal from a sensor/actuator loop. In this interpretation, \(\tau\) represents the effective loop latency or upstream feedback time, while \(C\) represents the effective feedback/coupling gain~\cite{Pyraginie}.

In \textbf{electronic} implementations (nonlinear RLC or Josephson-type circuits), a sinusoidally driven oscillator can be simultaneously perturbed by an upstream oscillator or a feedback loop. Again, $\tau$ captures the loop delay and $C$ the effective coupling strength~\cite{GronbechJensen1992,Kobayashi2022}. 
Finally, \textbf{delayed-feedback platforms} are widely used as laboratory substrates for emulating collective/network dynamics, where the delay and feedback gain play roles analogous to latency and coupling strength in extended systems~\cite{ZakharovaSemenov2025}. 
A \textbf{neuroscience-inspired} counterpart is a neural-mass or excitable unit receiving delayed synaptic input from an upstream population while also subjected to rhythmic stimulation; here $\tau$ models conduction/synaptic delays and $C$ synaptic efficacy~\cite{WangNeuro}.

For the delayed driver system, the initial condition must be specified as a history function on \( [-\tau,0] \). We adopt constant histories
\[
x_1(t)=u_0=1,\qquad \frac{dx_1}{dt}(t)=v_0=1,\qquad t\in[-\tau,0],
\]
as in~\cite{Coccolo_synch, Coccolo_synch2}, ensuring a well-defined and reproducible initial state for the delay differential equation. For the response system, which evolves in a finite-dimensional phase space, we set the initial conditions \(x_2(0)=0.5\) and \(\dot{x}_2(0)=0.5\). We have verified through additional simulations that the coupling--forcing induced coherence reported below persists qualitatively across a wide range of initial conditions. Thus, the results are not specific to the choices made here. We also note that the delayed oscillator evolves in an infinite-dimensional phase space, which is why history functions are required for the driver, whereas pointwise initial values suffice for the response. 

\begin{figure}[htbp]
  \centering
   \includegraphics[width=16.0cm,clip=true]{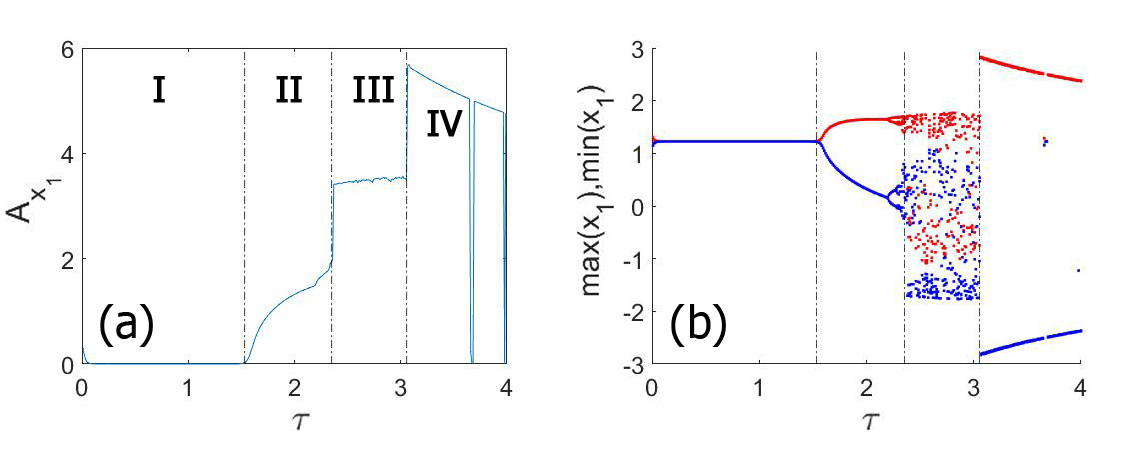}
\caption{Driver dynamics versus delay. (a) Steady-state peak-to-peak amplitude \(A_{x_1}\) of the delayed Duffing driver as a function of \(\tau\). (b) Maxima--minima diagram versus \(\tau\), showing the four qualitative regimes (I--IV) used throughout the paper.  The driver is unforced, with \(\mu=0.01\), \(\alpha=-1\), \(\gamma=-0.5\), and constant history functions \((u_0,v_0)=(1,1)\) for \(t\in[-\tau,0]\). In panel (b), maxima are shown in red and minima in blue.}
\label{fig:2}
\end{figure}

As reported in~\cite{Coccolo, Cantisan}, the unforced time-delayed Duffing
oscillator undergoes several bifurcations as the delay \(\tau\) is varied,
giving four dynamical regions, as summarized in Fig.~\ref{fig:2}. In that
figure, panel (a) shows the steady-state oscillation amplitude of the driver
\(A_{x_1}\), while panel (b) shows the maxima--minima diagram, obtained from
the extrema over the last five oscillation periods for each value of \(\tau\).
In Region~I (\(\tau<1.53\)), the dynamics converge to a fixed point. In
Region~II (\(1.53<\tau<2.35\)), sustained oscillations appear but remain
confined to one of the wells. In Region~III (\(2.35<\tau<3.05\)), the
amplitude jumps to values larger than the width of a single well, so that the
trajectories move from one well to the other and the dynamics become
aperiodic. Finally, in Region~IV (\(\tau>3.05\)), the trajectories are no
longer confined to either well and the oscillations become periodic again,
giving rise to a large limit cycle in phase space spanning both wells.
Regions~I and II provide regular driver inputs that can support organized
response behavior, whereas the inter-well aperiodic dynamics of Region~III and
the large-amplitude oscillations of Region~IV typically prevent, or mask, a
clearly localized coupling--forcing induced coherence region in the response.

Additional amplitude-based robustness scans with respect to a larger delay ($\tau=3.5$) are reported in Appendix~\ref{app:robustness} (Fig.~\ref{fig:robust_C_tau}).

 In the next section, we fix \(C=1.66\), following~\cite{Coccolo_synch, Coccolo_synch2}, where this coupling strength was shown to yield strong amplification in the response even without external forcing. This choice helps isolate and highlight the additional effects introduced by the independent periodic forcing. Later, we show that \(C=1.66\) lies in a region of high response amplitudes in the \(f\)--\(C\) parameter space, confirming its dynamical relevance for the phenomenon under investigation. While our primary focus is on the interplay between coupling and forcing, we have also examined the influence of key parameters. In particular, we explored the roles of \(C\) and \(\tau\), and investigated the effects of the forcing strength \(f\) and frequency \(\omega_2\) through bifurcation diagrams and amplitude maps. Variations of other parameters (e.g., \(\mu\) and the nonlinear coefficient) within reasonable ranges did not qualitatively alter the observed coherence phenomena, indicating that coupling--forcing induced coherence is robust across different parameter settings.

\subsection*{Coupling-induced equilibrium shift and mean-field correction}
\label{sec:xstar_shift}

\begin{figure}[htbp]
  \centering
   \includegraphics[width=16.0cm,clip=true]{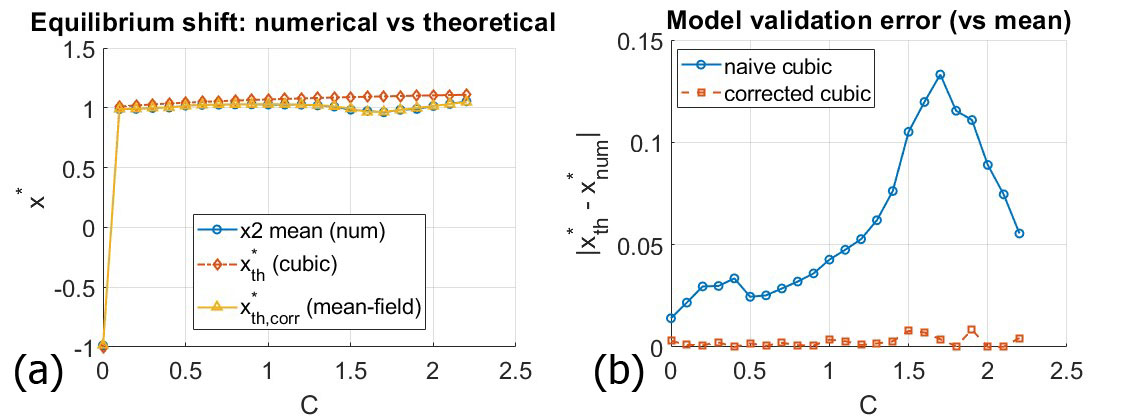}
\caption{Validation of the coupling-induced equilibrium shift for \(\tau=1\) and \(f=0\). (a) Numerical asymptotic mean \(x^\star_{\mathrm{num}}=\langle x_2\rangle\) versus coupling strength \(C\), compared with the naive cubic prediction (Eq.~(7)) and the mean-field corrected prediction (Eq.~(10)). (b) Absolute error \(|x^\star_{\mathrm{th}}-x^\star_{\mathrm{num}}|\) for the naive and corrected models, showing that accounting for finite residual oscillations (through \(\sigma_2\)) collapses the discrepancy across the full \(C\)-range. The remaining parameters are \(\mu=0.01\), \(\alpha=-1\), \(\gamma=-0.5\), with constant driver histories \((u_0,v_0)=(1,1)\) and response initial conditions \(x_2(0)=0.5\), \(\dot{x}_2(0)=0.5\).}
\label{fig:xstar_validation}
\end{figure}

In the parameter region where the delayed driver converges to a stationary state (Region~I), the coupling term in the response equation acts as an effective constant bias. Writing the response dynamics as
\begin{equation}
\ddot{x}_2+\mu \dot{x}_2+(C-1)x_2+x_2^3 = C x_1(t) + f\cos(\omega_2 t),
\label{eq:resp_rewritten}
\end{equation}
and using that $x_1(t)\approx \bar{x}_1$ asymptotically in Region~I, the biased equilibrium $x^\star$ (estimated with $f=0$) follows from the static balance
\begin{equation}
(x^\star)^3 + (C-1)x^\star - C\bar{x}_1 = 0.
\label{eq:cubic_naive}
\end{equation}
Figure~\ref{fig:xstar_validation}(a) confirms that the numerical asymptotic mean of the response, $x^\star_{\mathrm{num}}\equiv \langle x_2\rangle$, exhibits a clear coupling-induced shift consistent with the reduced description.

For intermediate coupling strengths, however, the response retains a finite residual oscillation in the steady window, so that the mean $\langle x_2\rangle$ need not satisfy Eq.~\eqref{eq:cubic_naive} exactly because $\langle x_2^3\rangle \neq \langle x_2\rangle^3$. To quantify this effect, we decompose $x_2(t)=m+\delta(t)$ with $m\equiv \langle x_2\rangle$ and $\langle \delta\rangle=0$. Expanding the cubic term yields
\begin{equation}
\langle x_2^3\rangle = m^3 + 3m\langle \delta^2\rangle + \langle \delta^3\rangle.
\label{eq:moment_expand}
\end{equation}
Assuming the residual oscillations are approximately symmetric over the averaging window (so that $\langle \delta^3\rangle$ is small), we obtain the leading-order moment-closure approximation
\begin{equation}
\langle x_2^3\rangle \approx m^3 + 3m\sigma_2^2,
\qquad \sigma_2^2\equiv \langle \delta^2\rangle,
\end{equation}
which leads to the corrected mean-field cubic
\begin{equation}
m^3 + \big[(C-1)+3\sigma_2^2\big]\,m - C\bar{x}_1 = 0.
\label{eq:cubic_corrected}
\end{equation}
The approximation $\langle \delta^3\rangle \approx 0$ corresponds to assuming that the residual oscillations around the mean are not strongly skewed over the averaging window, i.e., odd central moments are small compared with the variance term.
In practice, the excellent agreement of the corrected closure (Eq.~(10)) with the numerical means across the full explored coupling range indicates that any remaining contribution from $\langle \delta^3\rangle$ and higher-order cumulants is subdominant for the purposes of predicting the operating-point shift.
Residual discrepancies (visible only at the largest $C$ values) are therefore naturally attributable to neglected higher-order moments and finite-time averaging effects.

The corrected prediction $x^\star_{\mathrm{th,corr}}\equiv m$ agrees with the numerical mean across the entire $C$-range (Fig.~\ref{fig:xstar_validation}(a)), and reduces the model error by more than an order of magnitude relative to Eq.~\eqref{eq:cubic_naive} (Fig.~\ref{fig:xstar_validation}(b)). This validation supports the interpretation that coupling organizes the response by shifting its operating point, thereby modifying the local effective stiffness and the frequency-selective response discussed in the following sections.

\begin{figure}[!htbp]
  \centering
   \includegraphics[width=10.0cm,clip=true]{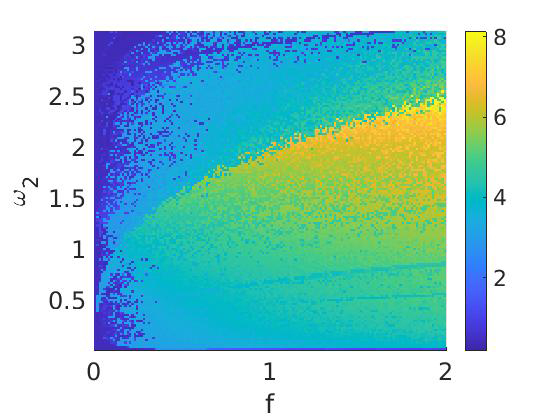}
\caption{Uncoupled response baseline (\(C=0\)). Steady-state peak-to-peak amplitude \(A_{x_2}\) as a function of forcing strength \(f\) and forcing frequency \(\omega_2\) for the response Duffing oscillator. This map provides the uncoupled baseline used for the cooperative-gain analysis; no coupling-dependent localized organization is present in this
case. The $(f,\omega_2)$ parameter plane is sampled on a $121 \times 121$ grid. The response parameters are \(C=0\), \(\mu=0.01\), and \(\alpha=-1\), with initial conditions \(x_2(0)=0.5\), \(\dot{x}_2(0)=0.5\).
}
   \label{fig:3}
\end{figure}

\begin{figure}[htbp]
  \centering
   \includegraphics[width=16.0cm,clip=true]{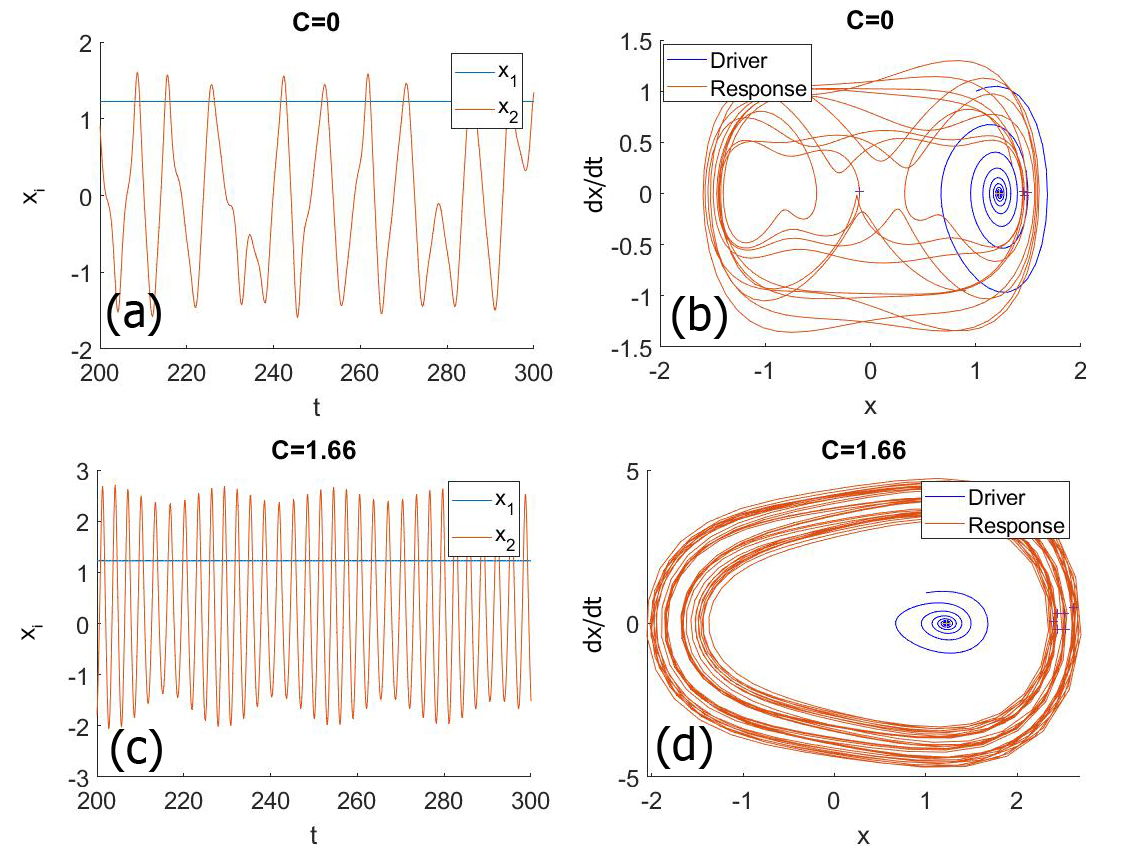}
\caption{Representative dynamics for \(\tau=1\) (\(C=0\) vs \(C=1.66\)). Panels (a)--(b): time series of the driver (blue) and response (red) for the same forcing parameters. Panels (c)--(d): corresponding phase portraits of the response. Large-amplitude response oscillations arise only when coupling and external forcing act simultaneously. In panels (b) and (d) the driver transient is included to illustrate convergence to the asymptotic state. The parameters are \(\tau=1\), \(\mu=0.01\), \(\alpha=-1\), \(\gamma=-0.5\), with \(C=0\) in panels (a)--(b) and \(C=1.66\) in panels (c)--(d). The forcing parameters are \(f=0.5\) and \(\omega_2=2\). The driver history is \((u_0,v_0)=(1,1)\), and the response initial conditions are \(x_2(0)=0.5\), \(\dot{x}_2(0)=0.5\).}
\label{fig:4}
\end{figure}

\begin{figure}[htbp]
  \centering
   \includegraphics[width=16.0cm,clip=true]{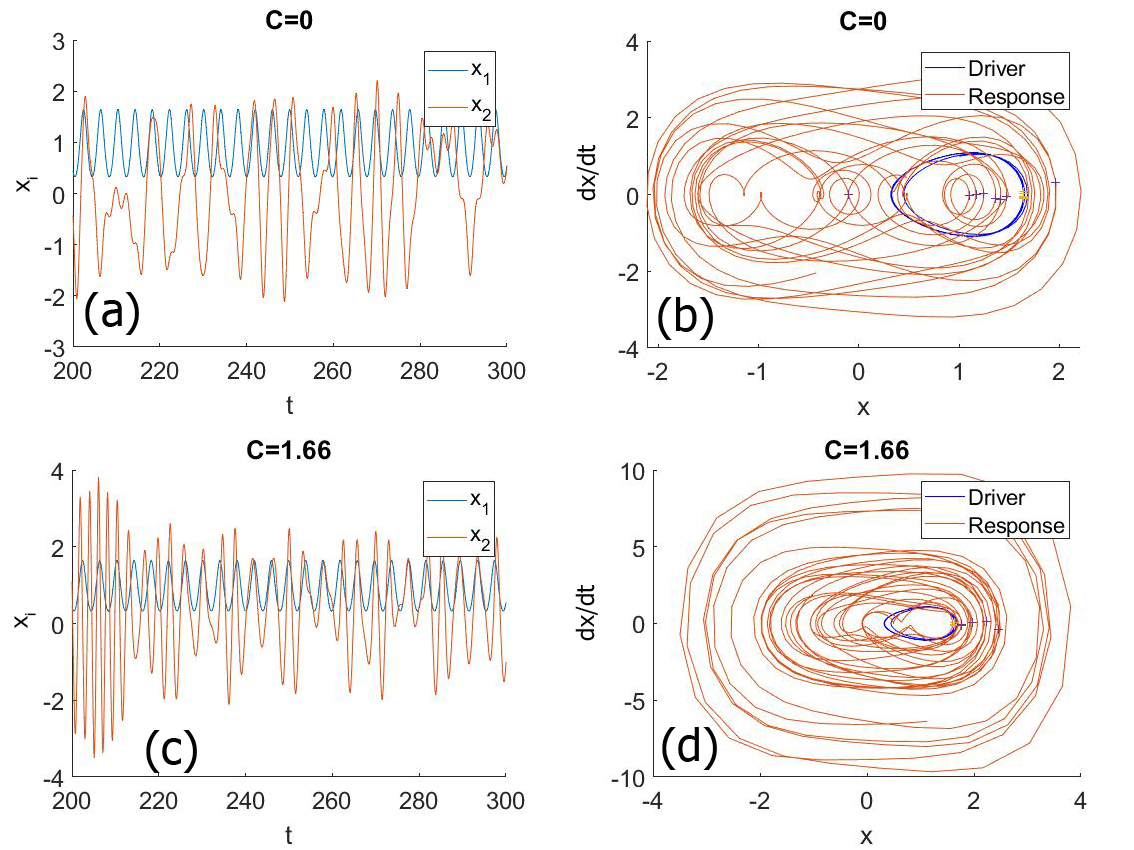}
 \caption{Representative dynamics for \(\tau=2\) (\(C=0\) vs \(C=1.66\)). Panels (a)--(b): time series of the driver (blue) and response (red) for the same forcing parameters. Panels (c)--(d): corresponding phase portraits of the response. Large-amplitude response oscillations arise only when coupling and external forcing act simultaneously. Panels (b) and (d) show the driver trajectory over the same window as the response, illustrating its asymptotic regime. The parameters are \(\tau=2\), \(\mu=0.01\), \(\alpha=-1\), \(\gamma=-0.5\), with \(C=0\) in panels (a)--(b) and \(C=1.66\) in panels (c)--(d). The forcing parameters are \(f=1.5\) and \(\omega_2=2.5\). The driver history is \((u_0,v_0)=(1,1)\), and the response initial conditions are \(x_2(0)=0.5\), \(\dot{x}_2(0)=0.5\).}
\label{fig:5}
\end{figure}

\section{Analysis of the coherence phenomenon}\label{Sec:III}

We now analyze the coherent organization induced by the combined action of the
coupling-mediated input and the external periodic forcing acting on the response system.
Before discussing the two-parameter maps, we stress that the term
``coupling--forcing induced coherence'' is not used here as a synonym for a
large-amplitude forced response. In the following, we distinguish between
(i) amplitude localization, meaning a restricted region of large steady-state
peak-to-peak response amplitude \(A_{x_2}\), and (ii) coupling--forcing induced
coherence, meaning a region where the response is both cooperatively enhanced
with respect to the uncoupled baseline and spectrally organized at the forcing
frequency. Thus, large amplitude alone is not sufficient to identify coherence.

\subsection{Diagnostic quantities and operational definition of coherence}
\label{ss:diagnostics}

Let \(A_{x_2}^{(C)}(f,\omega_2)\) denote the steady-state peak-to-peak
amplitude of the response for a given coupling \(C\), and
\(A_{x_2}^{(0)}(f,\omega_2)\) the corresponding amplitude with \(C=0\)
(uncoupled baseline). We define the \emph{cooperative gain}
\begin{equation}
G(f,\omega_2)
=
\frac{A_{x_2}^{(C)}(f,\omega_2)}
{A_{x_2}^{(0)}(f,\omega_2)+\varepsilon},
\qquad
\varepsilon=10^{-6},
\end{equation}
which measures the extra amplitude produced by coupling relative to the
uncoupled response at the same \((f,\omega_2)\). In the figures we often use
the decibel representation
\[
G_{\mathrm{dB}}(f,\omega_2)=20\log_{10}G(f,\omega_2).
\]

To quantify spectral organization we use the sharpness index~\cite{FreemanZhai2009}
\begin{equation}
\eta(f,\omega_2)
=
\frac{S(\omega_2;f,\omega_2)}
{\left\langle S(\omega;f,\omega_2)\right\rangle_{\omega\neq \omega_2}},
\end{equation}
where \(S(\omega;f,\omega_2)\) is the FFT amplitude of the steady-state
response signal. Thus, \(\eta>1\) indicates that the spectral component at the
forcing frequency is larger than the mean spectral background. For all spectra
we analyze the steady-state window of length \(T\), sampled with step
\(\Delta t\). Signals are demeaned and tapered with a Hanning window before
computing the FFT, and \(S(\omega)\) is reported without additional
normalization.

We also compute the dominant response frequency \(\omega_{2C}\) as the
frequency bin at which the FFT amplitude of the steady-state response is
maximal. This quantity is used as an additional frequency-domain diagnostic of
the organization of the response.

We label a point \((f,\omega_2)\) as \emph{coupling--forcing induced coherence}
when cooperative amplification and spectral concentration occur simultaneously,
namely
\begin{equation}\label{eq:G,eta}
G(f,\omega_2) \ge \kappa
\quad \text{and} \quad
\eta(f,\omega_2) > 1,
\end{equation}
with \(\kappa=1.5\) used for visualization. We verified that the qualitative
structure of the coherence region is robust for \(\kappa\in[1.3,1.7]\).
Accordingly, the coherence mask is not constructed from \(A_{x_2}\) alone.
This distinction is important because a large-amplitude forced response may
occur without satisfying the joint gain--sharpness criterion.

Throughout the paper, we therefore use three complementary notions.
\emph{Amplitude localization} refers to the appearance of a restricted
high-amplitude band in the response map \(A_{x_2}(f,\omega_2)\).
\emph{Spectral concentration} refers to the frequency-domain organization
quantified by \(\eta(f,\omega_2)\). Finally, the \emph{\cfcoh\ region}
(or \emph{coherence mask}) denotes the subset of parameter space satisfying
Eq.~\eqref{eq:G,eta}. When we refer to ``coherence'' below, we mean this
operational joint criterion, not amplitude enhancement alone.

\subsection{Baseline and representative dynamics}
\label{ss:baseline}

As a first step, Fig.~\ref{fig:3} shows the steady-state oscillation
amplitudes of the response when \(C=0\). This figure provides the uncoupled
baseline.
All amplitudes in Fig.~\ref{fig:3} are computed after discarding the transient
and therefore correspond to the asymptotic response reached from the prescribed
initial condition. We do not use this map as a global stability or
basin-of-attraction classification.
Regarding the nature of the patterns observed in Fig.~\ref{fig:3}, we
emphasize that this baseline map is intended to quantify the uncoupled amplitude
response, not to provide a complete dynamical classification of the uncoupled
Duffing oscillator. Depending on the forcing parameters, the uncoupled response
may include periodic, quasiperiodic, higher-period, or chaotic/irregular
oscillations. Therefore, the amplitude map should not be interpreted as a
one-to-one classification of the underlying attractor type.

In the present work, the role of Fig.~\ref{fig:3} is to provide the reference
amplitude \(A_{x_2}^{(0)}(f,\omega_2)\) used in the cooperative gain. A full
periodic/quasiperiodic/chaotic classification of the uncoupled response over
the entire \((f,\omega_2)\) plane would require additional diagnostics, such as
Lyapunov exponents, rotation numbers, or Poincaré sections, and is beyond the
scope of the present study.

Although the forced Duffing response can display finite, and even relatively
large, amplitudes in some parts of the \((f,\omega_2)\) plane, this baseline
does not exhibit the same coupling-dependent localized organization identified
below by the joint gain--sharpness criterion. Therefore, the relevant point is
not the complete absence of large oscillations in the uncoupled oscillator, but
the fact that the \cfcoh\ region arises only when the response is simultaneously
enhanced by coupling and spectrally concentrated at the forcing frequency.

This baseline is also important for separating the present mechanism from an
intrinsic resonance of the uncoupled response oscillator. If the uncoupled
response operates close to one of its own resonant regimes, then
\(A_{x_2}^{(0)}(f,\omega_2)\) may already be large. In that situation, a large
value of \(A_{x_2}^{(C)}\) does not by itself indicate coupling--forcing induced
coherence, because the cooperative gain
\(G=A_{x_2}^{(C)}/(A_{x_2}^{(0)}+\varepsilon)\) may remain small. Such points
are therefore excluded by the gain condition in Eq.~\eqref{eq:G,eta}.
Conversely, when coupling modifies the response through the driver-induced
bias or modulation, the coherent region is identified only if the coupled
response is enhanced relative to the uncoupled reference and is also
spectrally concentrated at the forcing frequency. Thus, possible intrinsic
resonances of the uncoupled response are not ignored; they are incorporated
into the reference state against which the cooperative effect is measured.

In Figs.~\ref{fig:4} and~\ref{fig:5}, we report representative results for
\(\tau=1\) and \(\tau=2\), respectively. These figures show that large and
organized response oscillations emerge when the response is simultaneously
influenced by the driver through the coupling term and directly driven by the
periodic forcing. In particular, the large responses in panels (c) and (d) of
Figs.~\ref{fig:4} and~\ref{fig:5} cannot be attributed only to the driver
(panels (a) and (b), blue curves) nor only to the uncoupled forced response
(panels (a) and (b), red curves). For comparison, Fig.~\ref{fig:3} reports the
uncoupled response amplitudes at the same level of description.

The results in Figs.~\ref{fig:4} and~\ref{fig:5} also highlight the role of
the coupling strength \(C\). When \(C=0\), the response remains comparatively
less organized for the selected parameters. By contrast, activating the
coupling (here \(C=1.66\)) produces a localized high-amplitude response and a
more structured phase portrait. This does not imply that all high-amplitude
states are coherent; rather, it motivates the use of the gain and spectral
diagnostics introduced above to determine which parts of parameter space
satisfy the operational coherence criterion.

This localized amplification is a cooperative effect of the external forcing
and the coupling-mediated perturbation: the coupling shifts or modulates the
response dynamics, while the periodic forcing selects the frequency content.
We use \(C=1.66\) as a representative intermediate-to-strong coupling value,
consistent with Refs.~\cite{Coccolo_synch, Coccolo_synch2}, to make this
organization clearly visible. Varying \(C\) in a neighborhood of \(1.66\)
preserves the same qualitative organization and mainly produces moderate
shifts in onset and bandwidth. Additional amplitude-based robustness scans with
respect to the coupling strength \(C\) are reported in Appendix~\ref{app:robustness}
(Fig.~\ref{fig:robust_C_tau}). Therefore, unless otherwise stated, we fix the
diffusive coupling to \(C=1.66\).

Although the underlying dynamics can be complex, the introduction of coupling
leads to markedly larger oscillation amplitudes for the selected representative
cases, particularly in the \(\dot{x}_2\) component. In Figs.~\ref{fig:4}(d)
and~\ref{fig:5}(d), the phase-space trajectory for \(C=1.66\) also exhibits a
more organized structure than in the uncoupled case. While a full
periodic/quasiperiodic/chaotic classification over the entire two-parameter
grid is beyond the scope of the present work, our identification of
coupling--forcing induced coherence is operational and is supported by
complementary diagnostics that directly quantify organization. Specifically,
inside the localized band the response displays large cooperative gain relative
to the uncoupled baseline, spectral concentration at the forcing frequency, and
enhanced temporal regularity, as quantified below by the coefficient of
variation of inter-peak intervals along representative parameter cuts.

Finally, Fig.~\ref{fig:4}(b) and (d) show the driver alongside the response
over the same time window to highlight the delayed modulation transmitted
through the coupling.

\begin{figure}[htbp]
  \centering
   \includegraphics[width=11.5cm,clip=true]{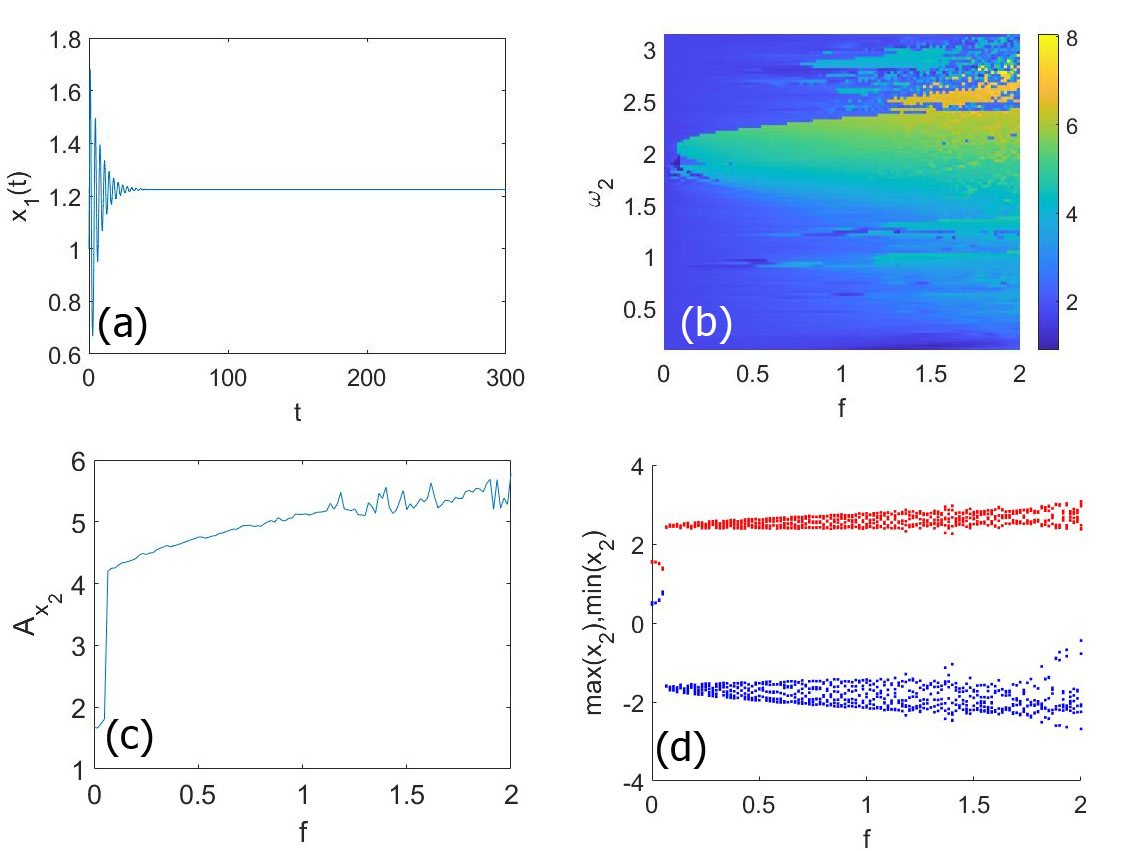}
\caption{Two-parameter organization for \(\tau=1\) at \(C=1.66\), with
\(F=0\), \(\mu=0.01\), \(\alpha=-1\), and \(\gamma=-0.5\).
(a) Driver behavior shown as representative oscillations.
(b) Response amplitude map \(A_{x_2}(f,\omega_2)\), showing a sharply localized
high-amplitude band.
(c) One-parameter cut: response amplitude \(A_{x_2}\) versus \(f\) at fixed
\(\omega_2=2\).
(d) Corresponding maxima--minima diagram versus \(f\) at \(\omega_2=2\).
The \((f,\omega_2)\) parameter plane is sampled on a \(121 \times 121\) grid. The driver history is \((u_0,v_0)=(1,1)\), and the response initial conditions are \(x_2(0)=0.5\), \(\dot{x}_2(0)=0.5\). In panel (d), maxima are shown in red and minima in blue.}
\label{fig:6a}
\end{figure}

\begin{figure}[htbp]
  \centering
   \includegraphics[width=12.5cm,clip=true]{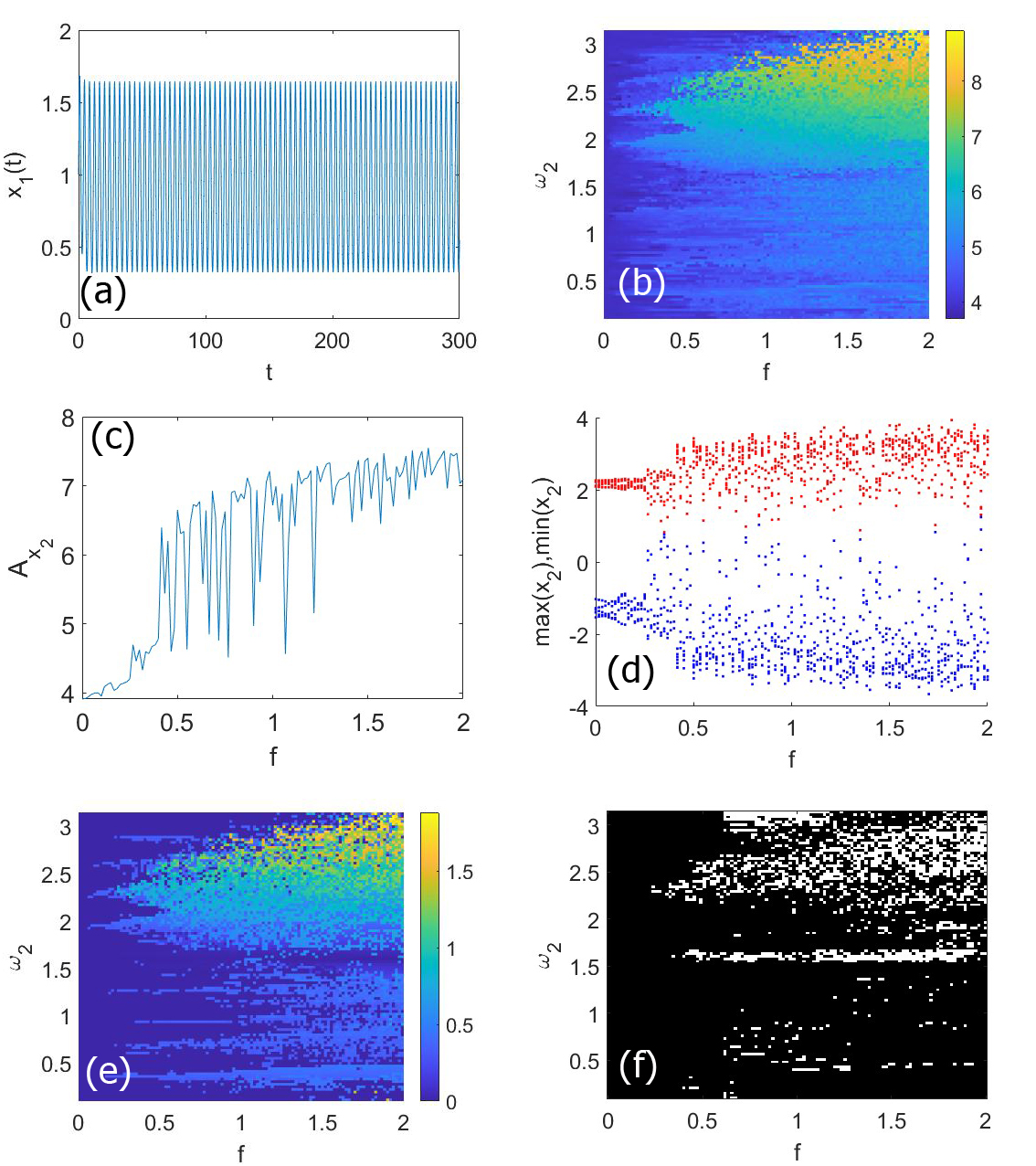}
\caption{Two-parameter organization for \(\tau=2\) at \(C=1.66\), with
\(F=0\), \(\mu=0.01\), \(\alpha=-1\), and \(\gamma=-0.5\).
(a) Driver oscillations.
(b) Response amplitude map \(A_{x_2}(f,\omega_2)\), showing a broader and more
diffuse localized band than for \(\tau=1\).
(c) One-parameter cut: response amplitude \(A_{x_2}\) versus \(f\) at fixed
\(\omega_2=2.5\).
(d) Corresponding maxima--minima diagram versus \(f\) at \(\omega_2=2.5\).
(e) Dominant response frequency \(\omega_{2C}\) from FFT.
(f) Spectral sharpness \(\eta\), with a white overlay indicating the
spectral-concentration region.
The \((f,\omega_2)\) parameter plane is sampled on a \(121\times121\) grid.
The driver history is \((u_0,v_0)=(1,1)\), and the response initial conditions are \(x_2(0)=0.5\), \(\dot{x}_2(0)=0.5\). In panel (d), maxima are shown in red and minima in blue.}
\label{fig:6b}
\end{figure}

\begin{figure}[htbp]
  \centering
   \includegraphics[width=11.5cm,clip=true]{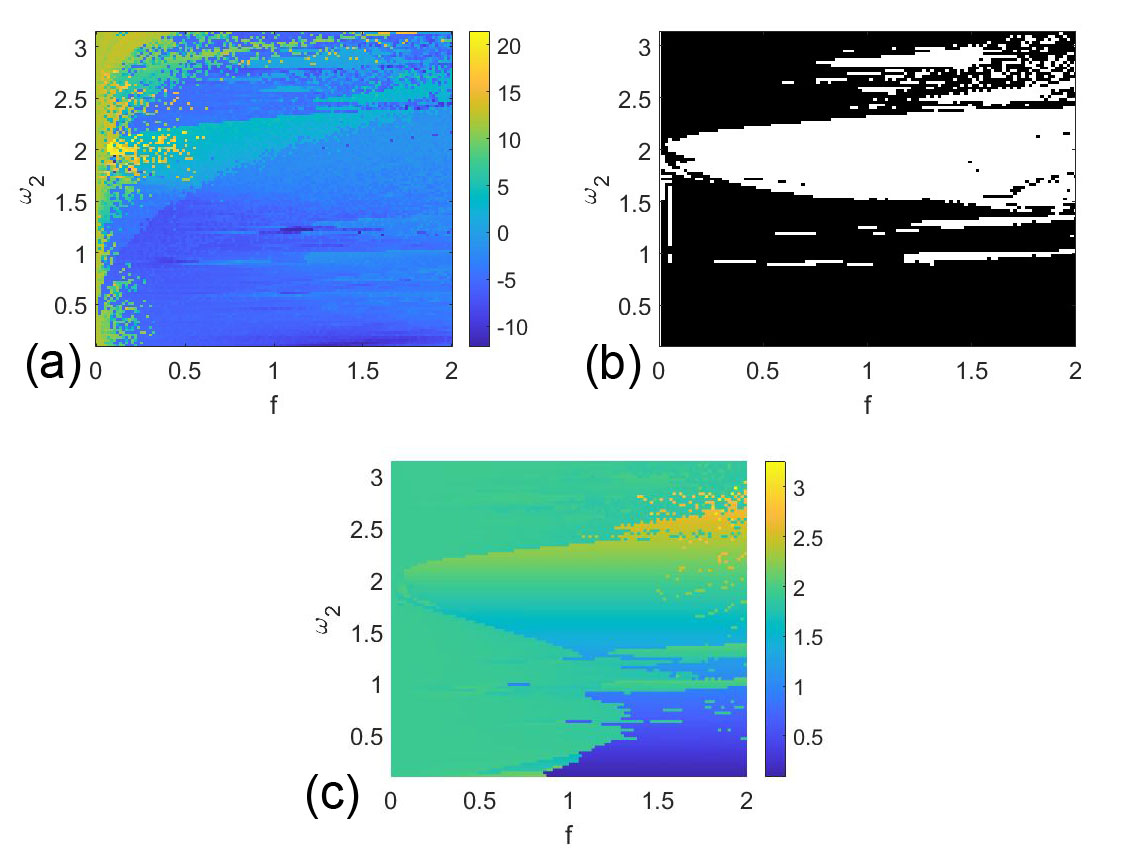}
\caption{Coherence diagnostics for \(\tau=1\), \(C=1.66\), \(F=0\),
\(\mu=0.01\), \(\alpha=-1\), and \(\gamma=-0.5\).
(a) Cooperative gain map
\(G_{\mathrm{dB}}(f,\omega_2)=20\log_{10}\!\big(A^{C=1.66}_{x_2}/A^{C=0}_{x_2}\big)\).
(b) Spectral sharpness \(\eta(f,\omega_2)\); the white overlay indicates the
region satisfying the spectral-concentration condition.
(c) Dominant response frequency \(\omega_{2C}(f,\omega_2)\) extracted from the
FFT on the steady-state window.
The \cfcoh\ region is defined by the joint condition \(G\ge\kappa\) and
\(\eta>1\) [Eq.~\eqref{eq:G,eta}], whereas amplitude localization refers only
to the high-\(A_{x_2}\) band in Fig.~\ref{fig:6a}(b). The
\((f,\omega_2)\) parameter plane is sampled on a \(121\times121\) grid.}
\label{fig:6a_bis}
\end{figure}

\begin{figure}[htbp]
  \centering
   \includegraphics[width=16.0cm,clip=true]{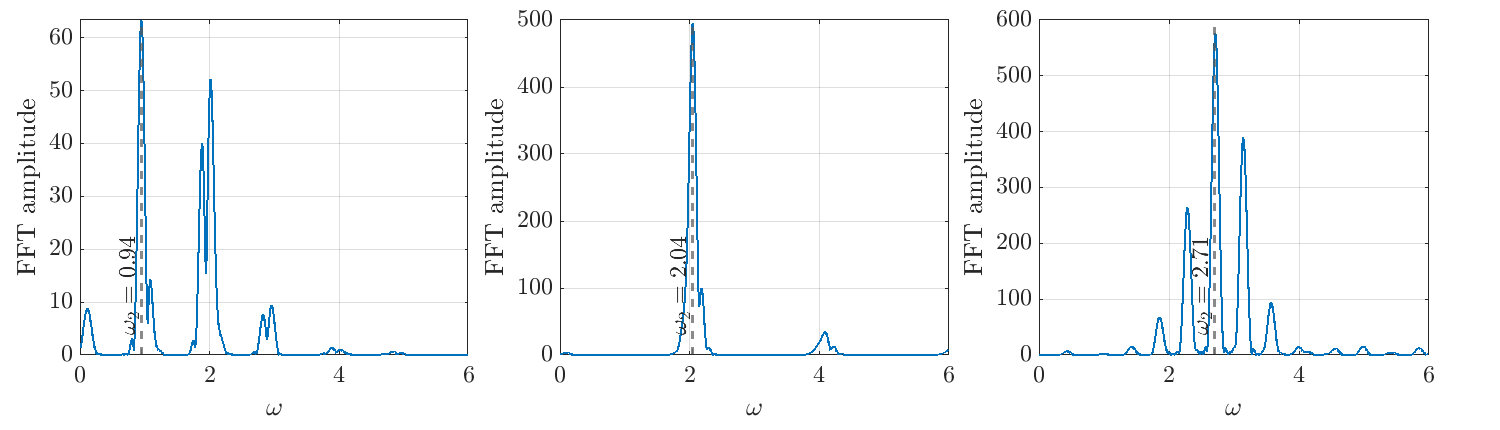}
\caption{
Representative steady-state spectra illustrating the operational distinction
between low-gain responses, coupling--forcing induced coherence, and
large-response points outside the coherence mask. The spectra are computed from
the response signal \(x_2(t)\) after removing the transient and using the same
FFT procedure employed for the sharpness index \(\eta\). The dashed vertical
line marks the forcing frequency \(\omega_2\) in each case.
Left: point outside the coherence mask with low cooperative gain,
\((f,\omega_2)=(0.681,0.941)\), \(A_{x_2}^{(C)}=1.138\),
\(A_{x_2}^{(0)}=4.450\), \(G=0.256\), and \(\eta=86.43\).
Middle: point inside the coherence mask,
\((f,\omega_2)=(0.107,2.044)\), \(A_{x_2}^{(C)}=4.406\),
\(A_{x_2}^{(0)}=0.401\), \(G=10.98\), and \(\eta=169.89\).
Right: large-response point outside the coherence mask,
\((f,\omega_2)=(1.943,2.706)\), \(A_{x_2}^{(C)}=7.385\),
\(A_{x_2}^{(0)}=5.043\), \(G=1.464\), and \(\eta=85.63\).
The comparison shows that neither a pronounced spectral peak nor a large
response amplitude alone is sufficient to identify coupling--forcing induced
coherence; only the joint criterion \(G\geq\kappa\) and \(\eta>1\) defines the
coherence mask. Parameters are \(\tau=1\), \(C=1.66\), \(F=0\),
\(\mu=0.01\), \(\alpha=-1\), and \(\gamma=-0.5\).
}
\label{fig:spectra}
\end{figure}

\begin{figure}[htbp]
  \centering
   \includegraphics[width=12.5cm,clip=true]{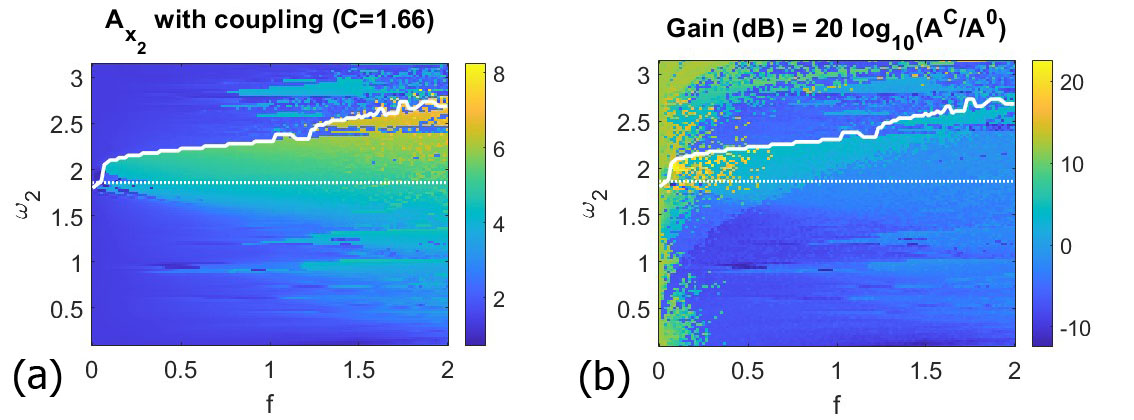}
\caption{Coupling-induced onset of the amplitude ridge predicted from the
biased equilibrium in Region~I. (a) Response amplitude map
\(A_{x_2}(\omega_2,f)\) for \(C=1.66\), with the extracted amplitude ridge
(solid line) and the theoretical onset frequency \(\omega_n\) from
Eq.~\eqref{eq:wn_theory} (dotted line). (b) Cooperative gain map
\(G_{\mathrm{dB}}(\omega_2,f)=20\log_{10}\!\big(A^{C}_{x_2}/A^{0}_{x_2}\big)\)
with the same overlays. The amplification band is anchored near
\(\omega_2\simeq\omega_n\) at small forcing and shifts upward with increasing
\(f\), consistently with nonlinear hardening about the biased operating point.
The \((f,\omega_2)\) parameter plane is sampled on a \(121 \times 121\) grid. The driver history is \((u_0,v_0)=(1,1)\), and the response initial conditions are \(x_2(0)=0.5\), \(\dot{x}_2(0)=0.5\). The coherence threshold is \(\kappa=1.5\).}
\label{fig:wn_ridge}
\end{figure}

\begin{figure}[htbp]
  \centering
   \includegraphics[width=16.0cm,clip=true]{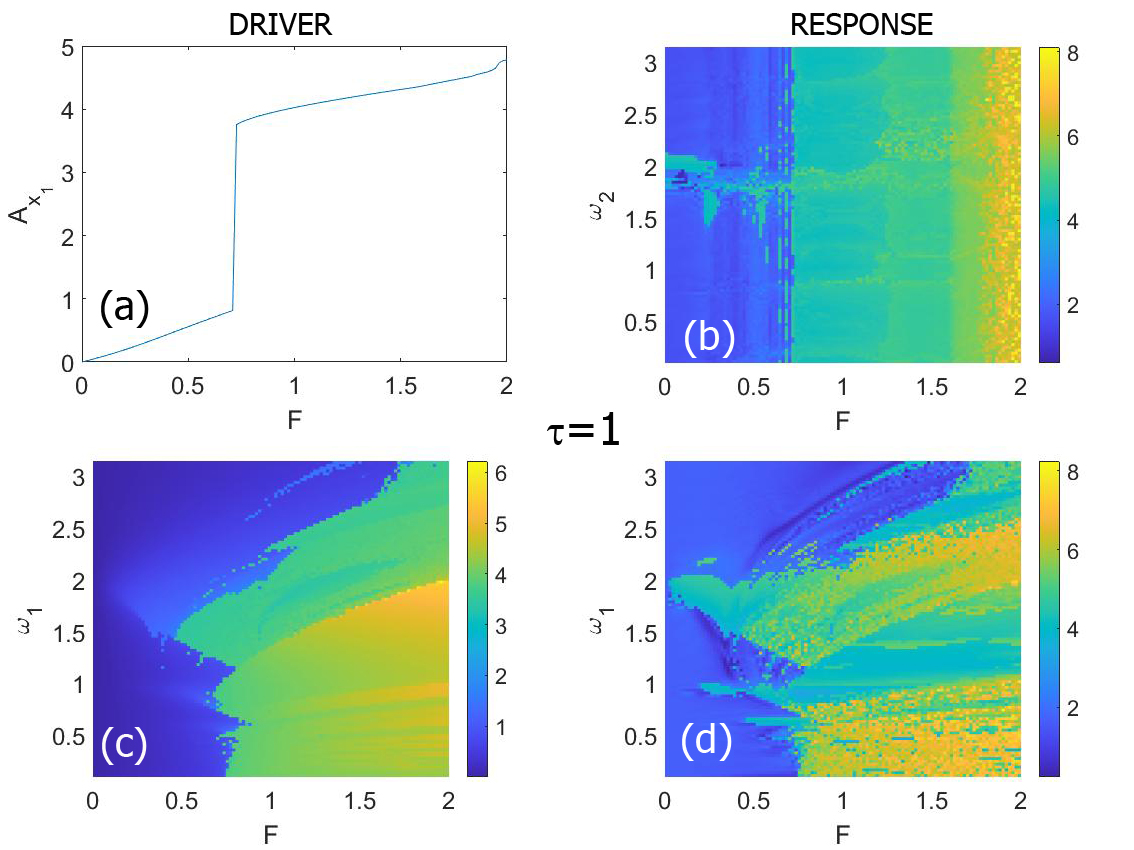}
\caption{Coexistence of coupling--forcing induced coherence and transmitted
resonance when both subsystems are externally forced (\(\tau=1\)).
(a) Driver amplitude \(A_{x_1}\) versus \(F\) for the parameters used in
panel (b).
(b) Response amplitude map in the \(F\)–\(\omega_2\) plane at fixed \(f\),
showing a broad high-amplitude region associated with transmitted resonance
at large \(F\), and a localized band near \(\omega_2\simeq 2\) at moderate
\(F\), consistent with coupling--forcing interaction in the response.
(c) Driver amplitude map in the \(F\)–\(\omega_1\) plane with \(f=0\).
(d) Corresponding response amplitude map, showing driver-forced transmission
effects together with a localized enhancement near \(\omega_1\simeq 2\) at
small \(F\). The parameter planes are sampled on a \(121 \times 121\) grid. The fixed parameters are \(\tau=1\), \(C=1.66\), \(\mu=0.01\), \(\alpha=-1\), and \(\gamma=-0.5\). In panels (a)--(b), \(\omega_1=1\) and \(f=0.1\) are fixed while \(F\) and \(\omega_2\) are varied. In panels (c)--(d), \(f=0\) and \(\omega_2=0.5\) are fixed while \(F\) and \(\omega_1\) are varied.}
\label{fig:7}
\end{figure}

\subsection{The effect of different \(\tau\) values}\label{ss:Cf}

To illustrate how the response amplitude depends on the forcing strength \(f\),
the forcing frequency \(\omega_2\), and the presence of coupling, we report the
two-parameter maps in Figs.~\ref{fig:6a} and~\ref{fig:6b} for \(\tau=1\) and
\(\tau=2\), respectively. A closer view of the underlying time series and phase
portraits for selected parameter values is provided in Figs.~\ref{fig:4}
and~\ref{fig:5}.

Despite belonging to different dynamical regions of the delayed driver, the two
maps share two robust features. First, a pronounced organization of the response
emerges around \(\omega_2 \approx 2\): for \(\tau\) in Regions~I--II, the
response is particularly prone to display coupling--forcing induced coherence
near \(\omega_2 \simeq 2\). Second, the localized band is a property of the
response under the combined action of direct forcing and coupling. Since, in the configuration of Eqs.~(4)--(5), the driver is not affected by the
response forcing parameters \((f,\omega_2)\), the localized structure observed
in the response cannot be attributed to a corresponding driver organization in
that parameter plane. Rather, it emerges from the response dynamics under the
combined action of the coupling-mediated input and the direct periodic forcing.

Compared with the uncoupled baseline (Fig.~\ref{fig:3}), panels (b) of
Figs.~\ref{fig:6a} and~\ref{fig:6b} show that coupling reshapes the forced
response into a localized high-amplitude region concentrated in a restricted
band of \(\omega_2\). However, as emphasized above, amplitude localization alone
is not sufficient to define coherence. Figure~\ref{fig:6a_bis} illustrates the
additional diagnostic ingredients entering the definition of the \cfcoh\ region:
the cooperative gain \(G\), shown in Fig.~\ref{fig:6a_bis}(a), which quantifies
the amplitude enhancement relative to the uncoupled baseline; the spectral
sharpness \(\eta\), shown in Fig.~\ref{fig:6a_bis}(b), which identifies
responses spectrally concentrated at the driving frequency; and the dominant
response frequency \(\omega_{2C}\), shown in Fig.~\ref{fig:6a_bis}(c). The
\cfcoh\ region is then defined by the joint satisfaction of
Eq.~\eqref{eq:G,eta}.

The comparison between the amplitude map and the diagnostic maps also shows why
the coherent region cannot be identified from amplitude alone. Some points may
have sizeable response amplitudes but fail the spectral-sharpness condition,
whereas other points may show spectral concentration without a large cooperative
gain. Only their intersection is labeled as coupling--forcing induced coherence.

To illustrate the meaning of the spectral sharpness criterion, Fig.~\ref{fig:spectra}
shows representative response spectra at three selected points: a low-gain point
outside the coherence mask, a point inside the coherence mask, and a large-response
point outside the mask. The comparison
confirms that the coherence mask is not defined by amplitude or spectral
concentration alone. A point may display a clear spectral component but have
low cooperative gain, and a large-response point may still remain outside the
mask if it does not satisfy the joint criterion. Only the simultaneous
satisfaction of \(G\geq\kappa\) and \(\eta>1\) identifies the
coupling--forcing induced coherence region.

A clear difference between \(\tau=1\) and \(\tau=2\) is that the localized
region in Fig.~\ref{fig:6b}(b) appears more diffuse than in
Fig.~\ref{fig:6a}(b). This is naturally related to the driver regime: for
\(\tau=1\) the driver converges to a stationary state, whereas for \(\tau=2\)
it approaches a periodic orbit. When the driver is stationary, the coupling
term \(C(x_1-x_2)\) contains a dominant quasi-constant component that biases
the response; when the driver is periodic, the coupling becomes more strongly
time-dependent and acts as a modulation, which tends to broaden the parameter
region where amplification is observed.

These differences are also evident in the one-parameter diagnostics shown in
Figs.~\ref{fig:6a}(c)--(d) and~\ref{fig:6b}(c)--(d), where the response
amplitude \(A_{x_2}\) and the maxima--minima diagrams are plotted versus \(f\)
at fixed forcing frequency. We use \(\omega_2=2\) for \(\tau=1\) and
\(\omega_2=2.5\) for \(\tau=2\), i.e., values that intersect the corresponding
organized bands. For \(\tau=1\), the response exhibits nearly steady
oscillations over a large part of the cut, whereas for \(\tau=2\) the maxima
and minima fluctuate more noticeably, consistently with a modulation-driven
regime. The frequency-domain organization is captured in
Figs.~\ref{fig:6a_bis}(c) and~\ref{fig:6b}(e), where \(\omega_{2C}\) denotes
the dominant response frequency extracted from the FFT on the steady-state
window. These panels show that the amplitude-localized band is also identifiable
in the frequency domain: within this band the response exhibits a well-defined
dominant frequency, although \(\omega_{2C}\) may vary across parameter space.
In addition, the FFT phase evaluated at the driving frequency \(\omega_2\)
varies smoothly across adjacent grid points inside the high-\(\eta\) region and
becomes irregular outside, consistently with enhanced phase consistency within
the \cfcoh\ region.

Finally, Fig.~\ref{fig:7} illustrates the case in which the driver is
also externally forced, so that coupling--forcing induced coherence and
transmitted resonance may coexist. Figure~\ref{fig:7}(a) shows the
driver response as the driver forcing amplitude \(F\) is varied, providing the
reference amplification transmitted through the coupling pathway. The response
map in Fig.~\ref{fig:7}(b) shows a broad high-amplitude region
associated with transmitted resonance at large driver forcing, together with a
localized band near \(\omega_2\simeq 2\) at moderate forcing. The complementary
maps in Figs.~\ref{fig:7}(c)--(d) show that driver-forced amplification
can be conveyed to the response through the coupling pathway, while localized
response organization may still occur in restricted parameter regions. Thus,
Fig.~\ref{fig:7} confirms that the two mechanisms are not mutually
exclusive: they can occupy different regions of parameter space or overlap when
both the driver and the response are externally forced.

\paragraph{Region dependence and bias-set onset frequency.}
We have also performed additional simulations by varying \(\tau\) continuously.
Within \(\tau\in[1,3]\), changes in the response amplitude occur gradually,
without abrupt transitions comparable to those of the driver in Fig.~\ref{fig:2}.
More pronounced changes appear when \(\tau\) crosses into regimes where the
driver dynamics becomes qualitatively different, such as chaotic or
large-amplitude motion. In these larger-delay regimes, we typically do not
observe a localized \cfcoh\ band of the type reported in Regions~I--II; instead,
the response is dominated by irregular or large driver-induced excursions.

In Region~I, the delayed driver converges to a stationary state
\(x_1(t)\to \bar{x}_1\), so the coupling term acts as an effective constant bias
in the response equation. This produces a coupling-dependent operating point
\(x^\star\) for the response, accurately captured by the static balance in
Eqs.~\eqref{eq:cubic_naive}--\eqref{eq:cubic_corrected}
(cf. Fig.~\ref{fig:xstar_validation}). Linearizing the response dynamics about
this biased equilibrium, \(x_2(t)=x^\star+\xi(t)\), yields the local
small-signal stiffness
\begin{equation}
k_{\mathrm{eff}}=(C-1)+3(x^\star)^2,
\end{equation}
and therefore the corresponding natural frequency
\begin{equation}
\omega_n=\sqrt{k_{\mathrm{eff}}}
         =\sqrt{(C-1)+3(x^\star)^2}.
\label{eq:wn_theory}
\end{equation}

Figure~\ref{fig:wn_ridge} compares the numerical organization of the response
with the onset frequency predicted from the biased equilibrium. Panel (a)
shows the numerical response amplitude map \(A_{x_2}(\omega_2,f)\). The
solid curve is the extracted amplitude ridge, obtained by selecting, for each
forcing level \(f\), the value of \(\omega_2\) that maximizes the response
amplitude inside the \cfcoh\ mask. Panel (b) shows the corresponding
cooperative gain map \(G_{\mathrm{dB}}(\omega_2,f)\), which indicates that the
same ridge is also associated with amplification relative to the uncoupled
baseline.

The dotted line in both panels represents the theoretical onset frequency
\(\omega_n\) obtained from Eq.~\eqref{eq:wn_theory}. The numerical ridge emerges
close to \(\omega_2\simeq\omega_n\) at small forcing, indicating that the
coupling-induced bias sets the frequency at which the organized response first
appears. As \(f\) increases, the ridge bends toward larger \(\omega_2\), which
is consistent with nonlinear hardening of the Duffing response around the
biased operating point. Thus, Fig.~\ref{fig:wn_ridge} shows not only the
theoretical onset estimate, but also the numerical amplitude and gain
structures that support the bias-set interpretation of the \cfcoh\ band.
Operationally, for each discrete forcing level \(f\), we compute
\(\omega_{2,\mathrm{ridge}}(f)\) as a direct \(\arg\max\) over the sampled
\(\omega_2\) values of \(A_{x_2}(\omega_2,f)\), restricted to points satisfying
the \cfcoh\ mask. No additional smoothing is applied; when multiple local
maxima occur within the mask, we select the global maximum, with ties broken by
choosing the smallest \(\omega_2\).

Outside Region~I, the driver is no longer stationary and the coupling term
becomes time-dependent, so the notion of a single constant operating point is
weakened. In Region~II, coupling--forcing induced coherence persists but is
better interpreted as modulation-driven organization rather than a bias-set
linear onset; accordingly, \(\omega_n\) provides only a qualitative anchor
there. In this region the driver converges to a periodic orbit, so the coupling
term \(C(x_1-x_2)\) acts as a time-dependent input rather than a quasi-constant
bias. From the response viewpoint, this produces an effective modulation of the
operating point and stiffness, which can broaden the parameter region where
amplification and spectral concentration occur. A natural interpretation is
therefore parametric/Floquet-type organization: linearizing the response about
the periodic driver-induced operating state yields a linear system with
periodic coefficients, for which selective growth of components near the forcing
frequency and its sidebands may occur. This mechanism would explain why the
coherent region in Region~II appears more diffuse than in Region~I. A
quantitative prediction of the Region~II boundaries would require a Floquet
analysis around the driver periodic orbit, which is beyond the scope of the present work.

\paragraph{Robustness to initial conditions.}
To assess the sensitivity of the \cfcoh\ identification to the choice of
initial histories and initial conditions, we repeated the full \((f,\omega_2)\)
scan on a reduced \(90\times 90\) grid for three distinct initial histories
(a baseline history plus two randomly perturbed histories). For each realization
we constructed the corresponding coherence mask using the same criteria adopted
throughout the paper, namely \(G\ge \kappa\), with \(\kappa=1.5\), and
\(\eta>1\), together with the same baseline-denominator regularization used to
avoid ill-conditioned gain ratios when the uncoupled amplitude is extremely
small. The resulting masks were identical across all three realizations.

To quantify this agreement we used the Jaccard overlap
(intersection-over-union) between two binary masks \(M_1\) and \(M_2\),
\begin{equation}
J(M_1,M_2)=\frac{|M_1\cap M_2|}{|M_1\cup M_2|},
\end{equation}
where \(|\cdot|\) denotes the number of grid points in the set. We obtained
\(J=1.000\) for both perturbed cases relative to the baseline, i.e., a perfect
overlap of the detected \cfcoh\ region within this scan and thresholding
procedure.

Since Duffing-type systems may exhibit coexisting attractors for some parameter
values, this test should not be interpreted as a proof that multistability is
absent everywhere inside the localized region. Rather, it shows that the
identified coherence mask is robust with respect to the tested perturbations of
the initial history and response initial conditions. A complete basin-of-attraction
analysis inside the localized band would require a dedicated multi-initial-condition
study, which is beyond the scope of the present work.

\paragraph{Additional temporal regularity diagnostic.}
As an independent check that the \cfcoh\ region corresponds to a more organized
response in time, and not only to large gain and spectral sharpness, we
evaluated a simple timing-regularity metric along two representative cuts,
\(f=0.5\) and \(f=1.5\). Specifically, in the steady-state window we detected
successive local maxima of \(x_2(t)\) at times \(\{t_k\}\) and computed the
inter-peak intervals \(\Delta t_k=t_{k+1}-t_k\). We then used the coefficient
of variation
\begin{equation}
\mathrm{CV}
=
\frac{\sigma(\Delta t_k)}{\langle \Delta t_k\rangle},
\end{equation}
as a dimensionless measure of peak-timing irregularity, with lower values
indicating more regular oscillations. Along both cuts, the CV remains low, of
order \(10^{-1}\), throughout the coherent-response region and exhibits
localized increases near transition zones, consistently with a degradation of
temporal regularity at the boundaries.

\subsection{The coupling constant effect}\label{ss:CCr}

We now examine how the coupling strength \(C\) shapes coupling--forcing induced coherence across delay regimes, and we comment on the recurring role of forcing frequencies near \(\omega_2\simeq 2\) observed in the previous subsection. In the transmitted-resonance scenario, \(\omega\simeq 2\) corresponds to a resonance-like frequency of the forced driver (for \(\tau=1\)) that is conveyed to the response through coupling. In the coupling--forcing induced coherence scenario (with \(F=0\)), \(\omega_2\simeq 2\) is instead the response forcing frequency around which the coupling-mediated bias/modulation and the periodic drive organize the response into a localized high-amplitude band. Importantly, in this latter case no analogous localized high-amplitude band is present in either subsystem alone; the localization emerges from their interaction.

\begin{figure}[htbp]
  \centering
   \includegraphics[width=16.0cm,clip=true]{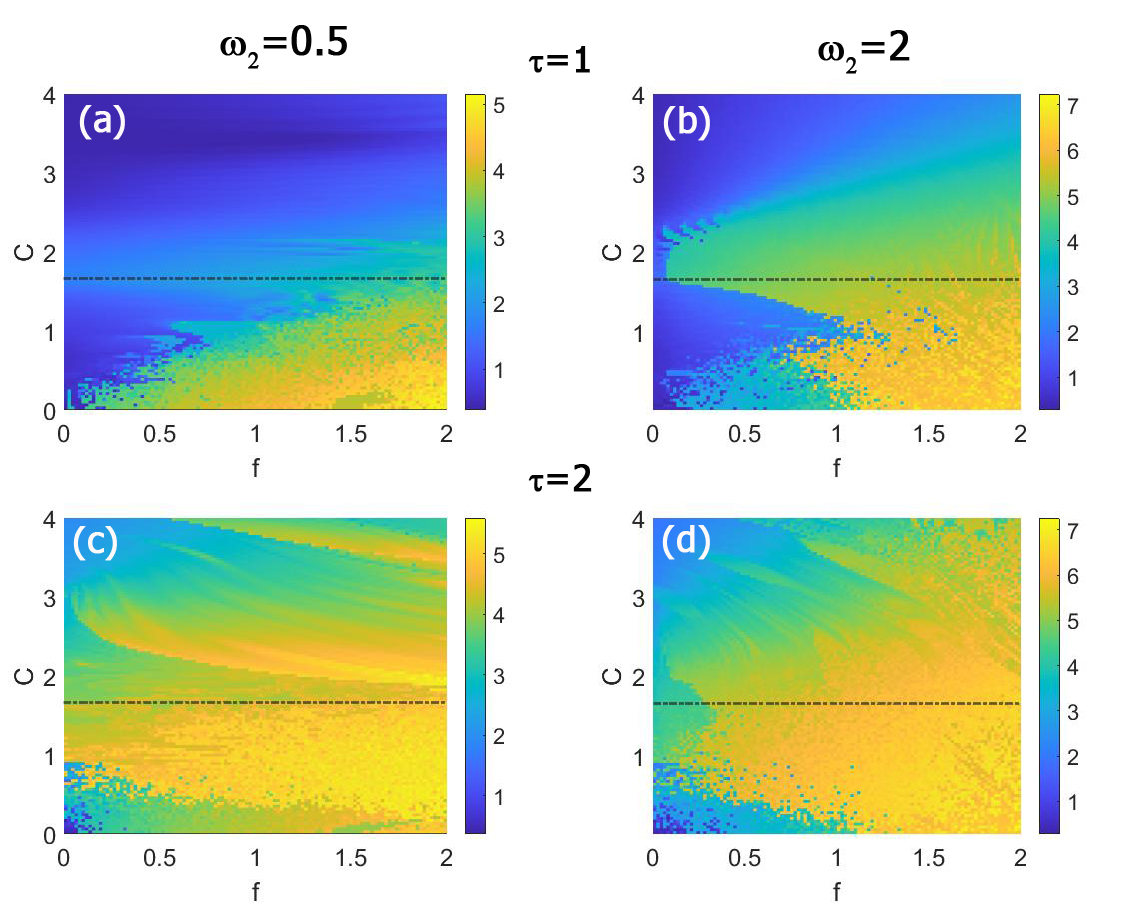}
\caption{Effect of coupling strength \(C\) on the response amplitude for
\(F=0\), \(\mu=0.01\), \(\alpha=-1\), and \(\gamma=-0.5\).
Response amplitude maps \(A_{x_2}(C,f)\) are shown for two forcing frequencies:
\(\omega_2=0.5\) (left column) and \(\omega_2=2\) (right column), and for two
delays \(\tau=1\) (top row) and \(\tau=2\) (bottom row). The black horizontal line
marks \(C=1.66\), used as the reference coupling in Sec.~III. The
\(\omega_2=2\) panels display larger and more structured response, consistent
with frequency-selective organization. The \((C,f)\) parameter plane is sampled
on a \(121 \times 121\) grid. The driver history is \((u_0,v_0)=(1,1)\), and the response initial conditions are \(x_2(0)=0.5\), \(\dot{x}_2(0)=0.5\).}
\label{fig:8}
\end{figure}

Figure~\ref{fig:8} reports response amplitude maps as functions of \((C,f)\) for different \(\tau\) regimes and for two representative forcing frequencies \(\omega_2\). The driver forcing is fixed at \(F=0\), as in Figs.~\ref{fig:6a} and~\ref{fig:6b}, to ensure that we are probing coupling--forcing induced coherence without contributions from transmitted resonance.

Several trends can be observed. First, the right-column panels (corresponding to \(\omega_2=2\)) display systematically larger response amplitudes than the left-column panels, consistent with the frequency-selective organization reported above. Second, for \(\tau\) in the regular-driver regimes (Regions~I--II), the high-amplitude response is not uniformly distributed in \((C,f)\) but organizes into structured regions. For example, in Fig.~\ref{fig:8}(b) the high-amplitude response appears more widespread and intermingled with low-amplitude regions than in Fig.~\ref{fig:8}(a), and a distinct wedge-like region of elevated amplitude is visible with its apex near \(C\approx 2\). By contrast, in the larger-delay regimes (panels (c) and (d)) the response displays broad areas of large amplitudes without a similarly sharp localized structure, consistent with the reduced ability of irregular/large driver dynamics to support organized response behavior.

Finally, Fig.~\ref{fig:8} also provides an a posteriori justification for using
\(C=1.66\) as a representative coupling value in the preceding sections: this
value lies in a region of consistently elevated response amplitudes across the
panels (black horizontal line). Overall, these results indicate a nontrivial
interplay between coupling strength \(C\) and forcing amplitude \(f\), whose
manifestation depends on the driver regime selected by \(\tau\).

\begin{figure}[htbp]
  \centering
   \includegraphics[width=16.0cm,clip=true]{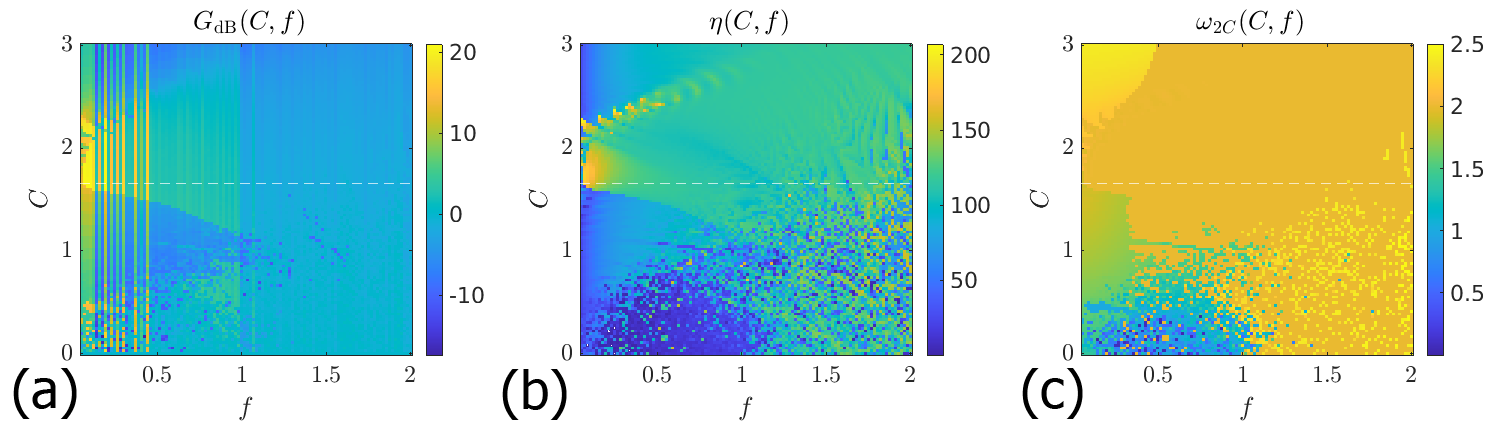}
\caption{Diagnostics in the \((C,f)\) plane for \(\tau=1\), \(\omega_2=2\),
\(F=0\), \(\mu=0.01\), \(\alpha=-1\), and \(\gamma=-0.5\).
(a) Cooperative gain map
\(G_{\mathrm{dB}}(C,f)=20\log_{10}\!\big(A^{(C)}_{x_2}/A^{(0)}_{x_2}\big)\).
(b) Spectral sharpness \(\eta(C,f)\), quantifying the concentration of the
response spectrum at the forcing frequency.
(c) Dominant response frequency \(\omega_{2C}(C,f)\) extracted from the FFT on
the steady-state window. The horizontal dashed line marks the reference value
\(C=1.66\). The maps show that the structured region observed in the amplitude
plot is associated not only with large response amplitude, but also with
cooperative gain and spectral organization around the forcing frequency,
consistent with the operational definition of coupling--forcing induced
coherence. The \((C,f)\) parameter plane is sampled on a \(110\times110\) grid.}
\label{fig:8bis}
\end{figure}

To further corroborate that the structures observed in the \((C,f)\) plane are
not merely large-amplitude regions, but are genuinely associated with
coupling--forcing induced coherence, we complement the amplitude maps of
Fig.~\ref{fig:8} with diagnostic maps in the same parameter plane.
Figure~\ref{fig:8bis} reports, for the representative case \(\tau=1\) and
\(\omega_2=2\), the cooperative gain \(G_{\mathrm{dB}}(C,f)\), the spectral
sharpness \(\eta(C,f)\), and the dominant response frequency
\(\omega_{2C}(C,f)\).

The gain and sharpness maps show that the structured response region in the
\((C,f)\) plane is accompanied by both cooperative amplification relative to
the uncoupled baseline and spectral concentration at the forcing frequency.
Moreover, the dominant-frequency map shows that this region remains organized
around the response forcing frequency. Therefore, the structures observed in
Fig.~\ref{fig:8} are not an artifact of plotting only \(A_{x_2}\), but remain
consistent with the operational definition of coupling--forcing induced
coherence. In particular, Fig.~\ref{fig:8bis} shows that this phenomenon is not
restricted to the single reference value \(C=1.66\), but persists over a
finite interval of coupling strengths.

We also checked the robustness of the main structures shown in
Figs.~\ref{fig:8} and~\ref{fig:8bis} with respect to changes in the initial
conditions and in the numerical integration parameters. Repeating representative
\((C,f)\) scans with perturbed driver histories and response initial conditions
preserved the main qualitative features of the maps, including the enhanced
response near \(\omega_2\simeq 2\) and the structured high-amplitude regions
around the reference coupling \(C=1.66\). We also verified that moderate changes
in the adaptive solver tolerances and maximum step size did not alter the
location of the dominant amplitude structures. These tests indicate that the
patterns in Figs.~\ref{fig:8} and~\ref{fig:8bis} are not artifacts of a
particular initial condition or of a specific numerical tolerance. As for the
coherence-mask robustness discussed above, this does not constitute a complete
basin-of-attraction analysis, but it supports the qualitative robustness of the
reported structures.

\section*{Reproducibility details}
All simulations were performed in MATLAB using \texttt{ddesd} with adaptive step size.
For the delayed driver, the history on $t\in[-\tau,0]$ was set to the constant functions $x_1(t)=u_0=1$ and $\dot{x}_1(t)=v_0=1$.
For the response system we used $x_2(0)=0.5$ and $\dot{x}_2(0)=0.5$.
Unless otherwise stated, parameters were fixed to $\mu=0.01$, $\alpha=-1$, $\gamma=-0.5$, and $C=1.66$, while $(f,\omega_2)$ were scanned on a uniform grid of size $N_f\times N_{\omega}$ over the ranges reported in the corresponding figure captions.

To compute all diagnostics, an initial transient interval was discarded and statistics were evaluated on a steady-state window corresponding to the last third of the total integration time $T=300$ time units.
For spectral quantities, the steady-state signal $x_2(t)$ was resampled on a uniform grid with sampling step $\Delta t=0.1$, demeaned, and tapered with a Hanning window prior to FFT computation.
The FFT was computed using a fixed transform length $N_{\mathrm{FFT}}=2^{18}$
(zero-padding the windowed steady-state signal when $N_{\mathrm{FFT}}>L$),
which sets the angular-frequency bin spacing
$\Delta\omega = 2\pi/(N_{\mathrm{FFT}}\Delta t)$.

The dominant response frequency $\omega_{2C}$ was extracted as the frequency bin maximizing the FFT amplitude within the resolved spectrum.
Response amplitude $A_{x_2}$ was computed as the peak-to-peak value over the steady-state window, and maxima--minima diagrams were constructed from the detected local extrema of $x_2(t)$.

Cooperative gain was defined as $G = A_{x_2}^{(C)}/(A_{x_2}^{(0)}+\varepsilon)$ with $\varepsilon=10^{-6}$, and points with extremely small uncoupled amplitude were excluded from gain-based masking to avoid ill-conditioned ratios.
Coupling--forcing induced coherence was then identified by the joint criterion $G\ge \kappa$ and $\eta>1$, with $\kappa=1.5$ used for visualization (robust for $\kappa\in[1.3,1.7]$).
 \section{Conclusions}\label{Sec:concl}

In this work, we analyzed the response of a nonlinear oscillator under two simultaneous external influences: a coupling-mediated input generated by a delayed driver system and an independent periodic forcing applied directly to the response. We showed that their cooperative action can induce a robust organization of the response dynamics, which we term \textit{coupling--forcing induced coherence}.

Coupling--forcing induced coherence appears as a localized high-amplitude band in parameter space that does not occur when either perturbation acts in isolation. In addition to amplitude localization, frequency-domain diagnostics reveal enhanced spectral concentration at the driving frequency within the same region, confirming that the effect involves both amplification and spectral organization. In this sense, the phenomenon is distinct from a purely resonance-based amplification picture and from transmitted resonance, since it requires the simultaneous presence of coupling and direct forcing.

We also examined the role of the delay parameter \(\tau\). Coherent organization emerges in the first two delay regions, where the driver dynamics remains regular (fixed point or periodic motion) and the coupling term can act effectively as a bias or a structured modulation of the response. For larger delays, where the driver exhibits irregular or very large-amplitude dynamics, organized response behavior is typically suppressed or visually masked by the strong driver-induced input. These results highlight that delay is not merely a technical modeling ingredient but a key factor selecting the driver regimes that can support coherence in the response.

Finally, we showed that coupling--forcing induced coherence may coexist with transmitted resonance when both subsystems are externally forced, and that the two mechanisms can overlap in parameter space. While separating their contributions point-by-point can be challenging in mixed regimes, the frequency-resolved maps provide a practical way to identify parameter regions where either mechanism dominates and where their combined action produces comparable amplification. Although demonstrated here for a delayed Duffing--Duffing configuration, the proposed framework and diagnostics are general and may be useful for interpreting analogous cooperative organization effects in other settings, including coupled mechanical oscillators, electronic circuits, and neuronal models.

The present study also has some limitations. First, the identification of
coupling--forcing induced coherence is operational and depends on the chosen
diagnostic thresholds, in particular the cooperative-gain threshold \(\kappa\)
and the spectral sharpness condition \(\eta>1\). Although we have verified that
the qualitative structure of the coherence region is robust for
\(\kappa\in[1.3,1.7]\), the exact boundaries of the mask should not be
interpreted as sharp bifurcation curves. Second, while additional simulations
indicate robustness with respect to representative changes in initial
conditions and numerical integration parameters, we have not performed a full
basin-of-attraction analysis over the entire parameter space; possible
multistability in Duffing-type dynamics therefore cannot be globally excluded.
Third, the theoretical interpretation based on the coupling-induced biased
equilibrium is most appropriate in Region~I, where the driver approaches a
stationary state. In Region~II, where the driver is periodic, the response is
better interpreted as modulation-driven, and a quantitative prediction of the
coherence boundaries would require a dedicated Floquet-type analysis. Finally,
the results have been obtained for a specific unidirectionally coupled
Duffing--Duffing configuration. Extensions to bidirectional coupling, larger
networks, noise, and experimental implementations remain outside the scope of this article.

Overall, our results motivate further exploration of coupling--forcing induced coherence in more complex networks, in the presence of heterogeneity and noise, and in experimentally realizable configurations where coupling-mediated inputs and external forcing naturally coexist.

\section{Acknowledgment}

This work was supported by the Spanish State Research Agency (AEI) and the European Regional Development Fund (ERDF,EU) under Project No. PID2023-148160NB-I00 (MCIN/AEI/10.13039/501100011033). It has also been supported by Rey Juan Carlos University, Spain, through the Programa Propio de Fomento y Desarrollo de la Investigación de la URJC under the project NABEH, funded by Financiación para proyectos Impulso (Project No. 2025/00014/047).

\appendix

\renewcommand{\thefigure}{A\arabic{figure}}
\setcounter{figure}{0}

\begin{figure}[htbp]
  \centering
  \includegraphics[width=16.0cm,clip=true]{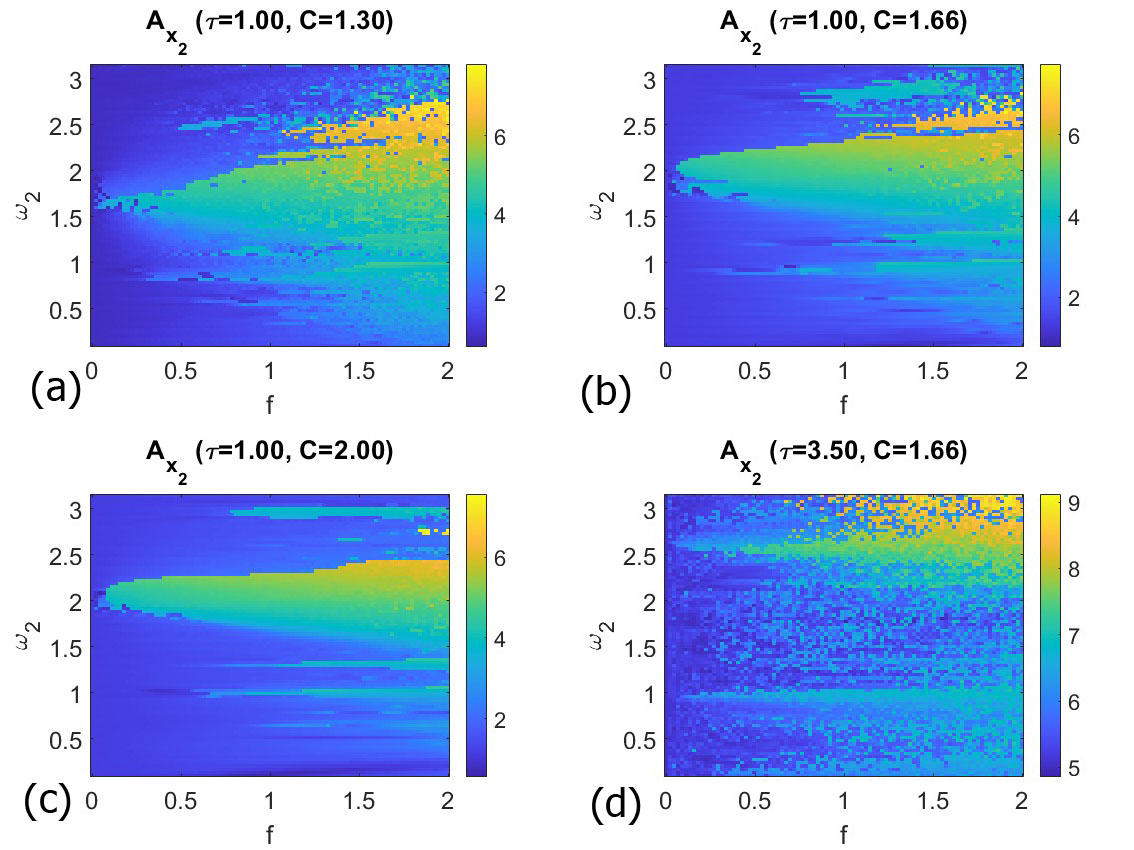}
  \caption{Amplitude-based robustness checks.
  (a--c) Steady-state response amplitude map $A_{x_2}(f,\omega_2)$ at $\tau=1$ for three coupling strengths: $C=1.3$ (a), $C=1.66$ (b), and $C=2.0$ (c), computed on a $90\times90$ grid in the $(f,\omega_2)$ plane.
  (d) Corresponding amplitude map for a larger delay $\tau=3.5$ at $C=1.66$, also computed on a $90\times90$ grid.
  Panels (a--c) use the same color scale for direct comparison across $C$, whereas panel (d) uses its own color scale (note different color limits) due to the markedly different overall amplitude level at larger delay.
  The coarse grid is used here for computational efficiency and to provide a qualitative robustness check; the main figures in the paper use a $121\times121$ grid. The fixed parameters are \(F=0\), \(\mu=0.01\), \(\alpha=-1\), and \(\gamma=-0.5\), with driver history \((u_0,v_0)=(1,1)\) and response initial conditions \(x_2(0)=0.5\), \(\dot{x}_2(0)=0.5\).}
  \label{fig:robust_C_tau}
\end{figure}

\section{Additional robustness checks}\label{app:robustness}

\paragraph{Amplitude-based robustness checks.}
Figure~\ref{fig:robust_C_tau} reports additional two-parameter scans based solely on the steady-state response amplitude $A_{x_2}(f,\omega_2)$.
For $\tau=1$ the localized high-amplitude organization persists when the coupling strength is varied over a representative range ($C=1.3,\,1.66,\,2.0$): the band remains anchored near $\omega_2\simeq 2$ at small $f$ and bends to higher $\omega_2$ as $f$ increases, consistent with a Duffing-type hardening about a coupling-biased operating point.
Changing $C$ primarily modulates the contrast and extent of this localized band, as expected from the scaling of the coupling-mediated input.
In contrast, for a larger delay ($\tau=3.5$ at $C=1.66$) the response becomes broadly amplified across the parameter plane and the localized band is no longer clearly isolated, consistent with the change of driver regime selected by larger delays and the resulting loss (or masking) of the mechanism that anchors localization in Regions~I--II.


\end{document}